\documentclass[aps,twocolumn,groupedaddress]{revtex4}
\usepackage[intlimits]{amsmath}
\usepackage{graphicx}
\usepackage[dvips,a4paper]{geometry}
\usepackage{amssymb}
\usepackage[latin1]{inputenc}
\usepackage[squaren]{SIunits}

\begin{document}
% You should use BibTeX and apsrev.bst for references
\bibliographystyle{apsrev}

\preprint{E-MRS Conference 2001, Symp. S}

\title{Template-Directed Self-Assembly of Buried Nanowires and the Pearling Instability}
% Optional argument for running titles on pages
%\title[]{}


\author{T. Müller, K.-H. Heinig, B. Schmidt}
\email[]{T.Mueller@fz-rossendorf.de}
\thanks{Tel. +49 351 260 3148; Fax. +49 351 260 3285}
%\thanks{}
\affiliation{Forschungszentrum Rossendorf, Institut f\"{u}r Ionenstrahlphysik \\und Materialforschung, PO-BOX 51 01 19, 01314 Dresden, Germany}


\date{\today}

\begin{abstract}
The fabrication of more and more miniaturized electronic and photonic devices
relies on new, ingenious methods for the fabrication of spatially controlled
nanostructures. Examples are electronic devices based on semiconducting
nanowires and photonic devices based on chains of metallic nanoclusters that
guide light by coupled surface plasmons. In this contribution a
template-directed ion beam synthesis of nanowires and regular nanocluster
chains will be presented. As templates, V-grooves etched in (001)Si and
subsequently oxidized are used. High fluence Ge$^{+}$ implantation is carried
out into the SiO$_{2}$ layer at \unit{70}{\kilo\electronvolt}. Thereby, the
implanted Ge enriches themselves in the V-groove bottom to a critical amount,
which may result in nanowire formation by nucleation, growth and coalescence
during subsequent thermal treatment. TEM investigations indicate the formation
of a nanowire buried in the SiO$_{2}$ at the V-groove bottom.

Kinetic lattice Monte Carlo simulations of the nanowire formation process were
performed in order to understand the phase separation mechanism and results
are compared to TEM images. Furthermore, it is shown that even ideal nanowires
show a instability and form during long-lasting annealing equal-spaced and
equal-sized nanoclusters (''nano\-cluster chains'') by self-organization.
\end{abstract}
%\keywords{}

\maketitle


\section{Introduction}

A wide field for novel applications opens at the bottom of the length scale
accessible by modern semiconductor preparation techniques. Synthesized
nanostructures show interesting physical properties being completely different
from the bulk material. Besides other applications, this could help to
miniaturize modern optical devices. Thus, having metallic nanoclusters
arranged in a regular chain, light can be transported by the coherent coupling
of surface-plasmon-polaritons \cite{Quinten:1998}. Along such a chain, light
can be guided around corners well below the classical diffraction limit,
allowing the integration of optical devices on the chip-level (Plasmonics)
\cite{Brongersma:2000}.

However, the precise synthesis of the nanostructures needed for such
applications remains a great challenge to current research. This becomes even
more important since classical structuring fails at this level. Here,
self-organization processes might be able to assemble structures with
nanometer dimensions. In particular, the ion beam synthesis (IBS) became
successful for fabrication of buried nano\-structures. Quantum dots were
formed under various conditions \cite{White:1997,Heinig:1998}. Using very high
ion fluences, buried layers were synthesized \cite{Li:1995}.

\begin{figure}[t]
\includegraphics[width=\columnwidth]{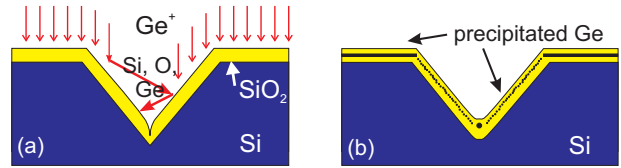}
\caption{Scheme of the ion beam synthesis of a buried Ge nanowire at
the bottom of an oxidized silicon V-groove. (a) Ge$^{+}$ ions are implanted at
high fluence into the V-groove. Atoms are sputtered from and are redeposited
onto the sidewalls leading to a mass transport towards the V-groove bottom.
During annealing (b) in an inert ambient a nanowire may form at the V-groove bottom.}
\label{fig:Scheme}
\end{figure}

In order to obtain spatially well-defined nanostructures, taming of
self-organization of pattern formation is an attractive approach. In the
present contribution, this is achieved by oxide covered Si V-grooves as
templates, which guide the self-assembly of buried nanowires from implanted Ge
atoms during thermal treatment. This scheme is depicted in
Fig.~\ref{fig:Scheme}. By a similar approach Si nanowires were recently
fabricated by O$^{+}$ implantation into Si V-grooves \cite{Ishikawa:1999}.
However, the method of ref. \cite{Ishikawa:1999} is restricted to Si wires and
is hard to control. Here, Ge was implanted as it is a well-known material for
nanocluster synthesis. However, also other ions can be easily applied, e.g. Au
for plasmonic devices. Different stages of the fabrication process were
monitored by cross sectional Transmission Electron Microscopy (X-TEM) investigations.

Theoretical studies are presented to understand details of the phase separation mechanism. On the atomic scale, within kinetic 3D lattice Monte Carlo simulations (KLMC), the diffusion of Ge atoms, their nucleation to clusters, cluster ripening and coalescence are tracked. For long lasting thermal treatment such simulations also describe the pearling instability of wires. A nanowire tends to decay into droplets driven by the minimization of interfacial energy and highly regular chains of equal-sized and equal-spaced nanoclusters form.

\section{Experiments}

\begin{figure}
\includegraphics[width=\columnwidth]{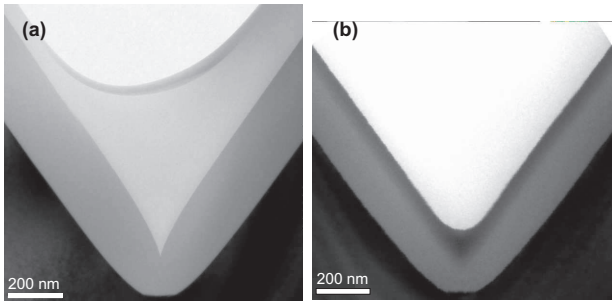}
\caption{XTEM images of (a)\ an as-oxidized and (b) an as-implanted
sample (\unit{70}{\kilo\electronvolt}, \unit{1\times10^{17}}{Ge^{+}\centi\meter^{-2}}). The Ge content in the SiO$_{2}$ appears as dark shadow in the TEM image (b).}
\label{fig:Both}
\end{figure}

V-grooves aligned along the [110] direction on (001)Si wafers were prepared by photolithographic masking and anisotropic etching with $30\%$ KOH at \unit{80}{\celsius}. This self-adjusting process results in atomically smooth (111) crystal facets forming the sidewalls of the V-grooves with an angle of \unit{54.7}{\degree} to the plane wafer surface. V-grooves of \unit{4}{\micro\meter} width and \unit{5}{\milli\meter} length were studied. They were oxidized in dry O$_{2}$ at \unit{1000}{\celsius}, leading to \unit{200}{\nano\meter} thick SiO$_{2}$ on the (001)Si surface and \unit{220}{\nano\meter} thick SiO$_{2}$ on the (111) sidewalls of the V-groove. A sharp surface depression forms during oxidation at the V-groove bottom, which can be seen in X-TEM images (Fig.~\ref{fig:Both}~a). The SiO$_{2}$ covered V-grooves were implanted at room temperature with \unit{1\times10^{17}}{Ge^{+}\centi\meter^{-2}} at \unit{70}{\kilo\electronvolt} (flux density \unit{3\times10^{12}}{\centi\meter^{-2}\second^{-1}}). The content of Ge in the SiO$_{2}$ appears thereby as dark shadow in the TEM image (Fig.~\ref{fig:Both}~b). During implantation, Ge is enriched in the V-groove bottom compared to the sidewalls due to (i) the reduced ion density per surface area at the sidewalls caused by the inclination and (ii) due to ion erosion of the sidewalls. Sputtered atoms including Ge are redeposited at the opposite sidewall and refill the sharp kink at the as-oxidized V-groove bottom. The V-groove surface evolution under ion bombardment was recently modeled in a continuum framework \cite{Mueller:2000} in nice agreement with V-grooves shapes observed in TEM images also proving a Ge accumulation of $30\%$ in the V-groove bottom.

\begin{figure}
\includegraphics[width=\columnwidth]{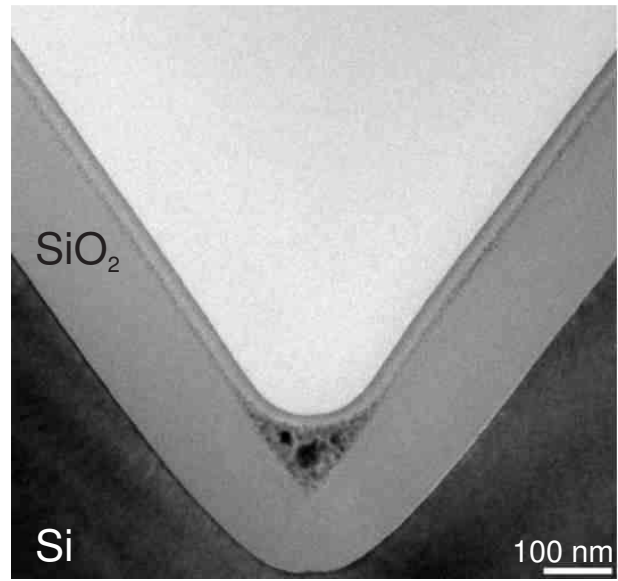}
\caption{XTEM image of an annealed sample. The large dark spot in the V-groove bottom appears to be a cross section of a Ge nanowire with \unit{35}{\nano\meter} diameter.}
\label{fig:QWR-TEM}
\end{figure}

\begin{figure}
\includegraphics[width=0.7\columnwidth]{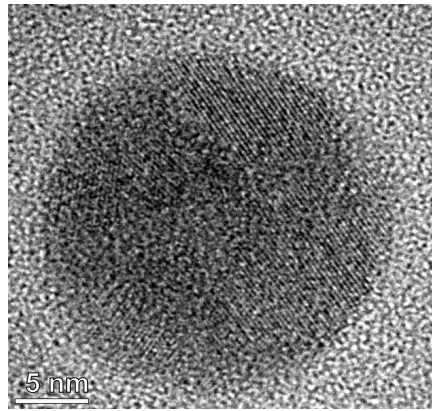}
\caption{High resolution TEM image of a Ge cluster in an annealed sample showing lattice fringes, which proves the crystallinity of the precipitates.}
\label{fig:cluster}
\end{figure}

Subsequent furnace annealing in N$_{2}$ ambient was performed at \unit{950}{\celsius} for $10$ minutes. Thereby, the small amount of implanted Ge is extremely sensitive to moisture, which may result in the oxidation of Ge. Therefore, (i)
the standard wet cleaning step in H$_{2}$O$_{2}$/H$_{2}$SO$_{4}$ prior to any thermal processing was skipped and (ii) the samples were transferred into the cold furnace and then purged in cold N$_{2}$ in order to reduce the moisture
intake. Fig.~\ref{fig:QWR-TEM} shows an XTEM image of an annealed sample. Buried Ge clusters are present in the sidewalls and the bottom. The dark dot, visible in the V-groove bottom, is a cross-section of the expected nanowire
with a diameter of \unit{35}{\nano\meter}. All Ge cluster are crystalline, which is confirmed by a high resolution TEM
image, Fig. \ref{fig:cluster}, of a Ge cluster showing lattice fringes.

The continuity of the wire over long distances remains to be proven. Several XTEM images taken from different cross-sections show always the large black dot with identical diameter and position in the V-groove bottom, which is an
indication for a continuous nanowire.

\section{Kinetic Monte-Carlo Simulations}

KLMC simulations were performed in order to understand the wire formation from dispensed Ge atoms. The interactions between impurity atoms (Ge) were simulated by a Cellular Automaton approach. Gauging the bond strength $E_{B}$
of Ge atoms, their \emph{effective} nearest-neighbor interactions within the host matrix, the SiO$_{2}$, must be considered. Here, the most simple implementation is used, i.e. the interaction is assumed to be configuration
independent. The binding energy $nE_{B}$ of a Ge atoms is just proportional to the number $n$ of nearest-neighbor sites occupied by Ge and can be described by the Ising model. Additionally, the diffusion of Ge monomers in the host
matrix is thermally activated with an activation energy $E_{A}$.

Then, the transition from the initially out-of-equilibrium state towards thermodynamical equilibrium is described by the importance sampling of configurations according the Metropolis algorithm \cite{Metropolis:1953}. A sequence of states is produced representing the time evolution, where each state is generated from its predecessor by a jump of a randomly chosen Ge atom to an arbitrary empty nearest-neighbor site with the transition probability
\[
P_{if}\sim\left\{
\begin{array}
[c]{cc}%
\exp\left\{  -\frac{E_{A}}{k_{B}T}\right\}  , & \; n_{f}\geq n_{i,}\\
\exp\left\{  -\left(  n_{i}-n_{f}\right)  \frac{E_{B}}{k_{B}T}-\frac{E_{A}%
}{k_{B}T}\right\}  , & \; n_{f}<n_{i}.
\end{array}
\right.  .
\]
During one Monte Carlo step (MCS) each atom performs statistically one attempt to jump from its initial site $i$ to a final site $f$.\ Thereby, the activation energy for diffusion $E_{A}$ is assumed to be the same for each
possible transition, i.e. for interface and bulk diffusion. Thus, it is possible to renormalize the transition probability $P_{if}$ such that each diffusional jump is successful $(E_{A}\rightarrow0)$, which only scales the
time of a MCS thereby improving the simulation speed 
\[
P_{if}\sim\left\{
\begin{array}
[c]{cc}%
1, & \; n_{f}\geq n_{i,}\\
\exp\left\{  -\left(  n_{i}-n_{f}\right)  \frac{E_{B}}{k_{B}T}\right\}  , &
\; n_{f}<n_{i}.
\end{array}
\right.  .
\]
This approach has been proven useful for the description of precipitation after ion implantation \cite{Strobel:1998}, and the evolution of fcc surfaces under ion erosion \cite{Strobel:2001}.

\begin{figure}
\includegraphics[width=\columnwidth]{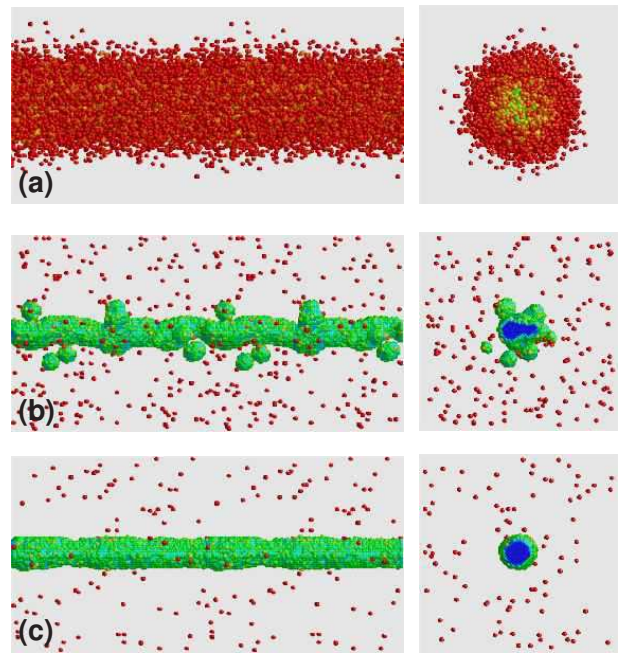}
\caption{Kinetic lattice Monte Carlo simulation of the nanowire formation from an initially cylindrical Gaussian distribution of impurity atoms with a peak concentration of $31\%$ and \unit{4}{\nano\meter} standard deviation. Shown are (a) the initial, (b) an intermediate ($102$ kMCS), and (c) the final ($490$ kMCS) state. The final wire diameter is
approximately \unit{5}{\nano\meter}.}
\label{fig:KMC}
\end{figure}

For the KLMC simulations of the nanowire formation a simplified initial distribution of Ge atoms was assumed. A cylindrical profile with a Gaussian cross section aligned to the [100] direction of the simulation lattice was
used (Fig.~\ref{fig:KMC} a) with a peak concentration of $31\%$ and\unit{4}{\nano\meter} standard deviation. $95\%$ of the implanted atoms lay in a cylinder of \unit{16\{\nano\meter} diameter, which resembles the Ge-rich region at the V-groove bottom, see Fig.~\ref{fig:Both}b).

\begin{figure*}
\includegraphics[width=0.8\textwidth]{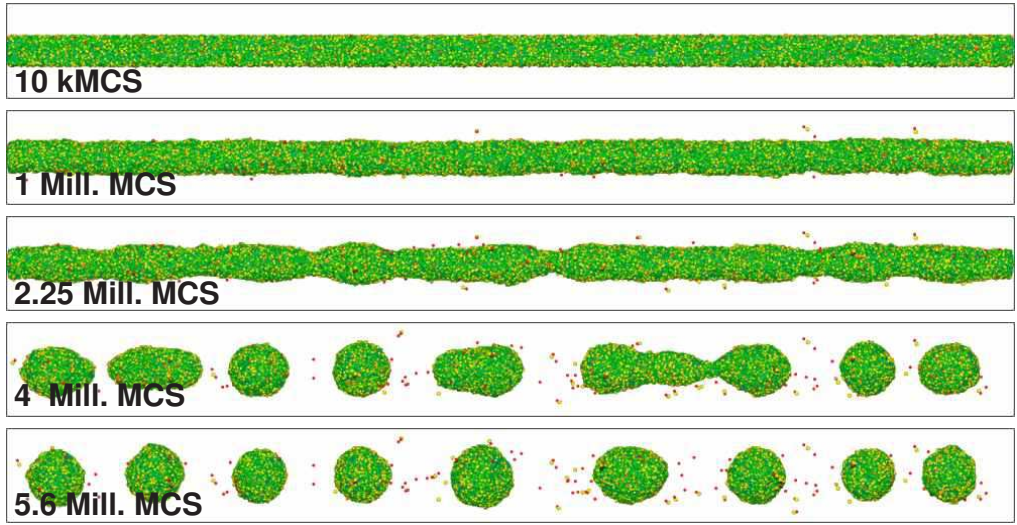}
\caption{Kinetic lattice Monte Carlo simulation of the pearling instability of a nanowire with \unit{8.6}{\nano\meter}
diameter. The wire decay proceeds via interface diffusion and a chain of nanoclusters with an average diameter of \unit{16.4}{\nano\meter} is formed.}
\label{fig:fragmentation}
\end{figure*}

During annealing, Ge nanocluster nucleate from the supersaturation and grow in time. They mostly have nucleated at the position of highest supersaturation, which is the profile center. As they grow, they start to touch each other and
coalesce to extended structures. A continuous wire forms if this coalescence proceeds further, while the surrounding nanoclusters evaporate and Ge condenses onto the wire-like structure, see Fig.~\ref{fig:KMC}b,c). The complete coalescence occurs only, if the initial profile is not too broad and the Ge concentration not too low, i.e.~if peak concentration of the initial Ge profile is larger than 25\% (at the given profile width). Otherwise, the clusters remain isolated and randomly distributed in size and space.

Comparing experiment and KLMC simulations, it is obvious that a buried nanowire could form at the V-groove bottom at the chosen Ge$^{+}$ ion fluence. As seen from Fig. \ref{fig:QWR-TEM}, the annealing has been interrupted before
all surrounding nanoclusters have disappeared, which compares to Fig.~\ref{fig:KMC}b).

It should be noted that the KLMC simulation uses a lattice. Thus, a wire formed by coalescence will be single crystalline, whereas an experimentally fabricated wire is expected to be polycrystalline. Therefore, the influence of
grain boundaries is not taken into account in the present study.

\section{Pearling Instability}

For long-lasting annealing, theory predicts that a continuous wire decays. Thus, a wire is an unstable configuration. Its decay is driven by the minimization of the interfacial energy. Harmonic perturbations of the wire radius lower its interfacial energy if their wavelength is larger than the circumference of the wire, $\lambda>2\pi R_{0}$. This result has been first found by Lord Rayleigh \cite{Rayleigh:1879} in 1879 for inviscid liquids. Linear stability analysis also revealed a fastest growing mode $\lambda_{m}$ of radius fluctuations during the initial stage of the fragmentation. Thus, perturbations with wavelength $\lambda_{m}$ grow at the fastest rate in time and are expected to dominate fragmentation. The value of $\lambda_{m}$ depends only on the specific mass-transport mechanism and not on material parameters or temperature. For liquid ''wires'', which changes their shape by hydrodynamic flow, a value of $\lambda_{m}\cong9.02\cdot R_{0}$ was predicted \cite{Rayleigh:1879}, which is also valid for internal volume diffusion as shown by Nichols and Mullins \cite{Nichols:1965}. For surface diffusion, it holds
\[
\lambda_{m}=\sqrt{2}2\pi R_{0}\cong8.89\cdot R_{0},
\]
while for external volume diffusion the value a $12.96\cdot R_{0}$ was obtained for $\lambda_{m}$ \cite{Nichols:1965}. With the result of the stability analysis an average droplet diameter $d_{cluster}$ can be given, if one assumes that the volume between two consecutive radii minima forms a droplet. For surface diffusion this gives
\[
d_{cluster}=\sqrt[3]{6\lambda_{m}R_{0}}=3.\,78\cdot R_{0}.
\]
In order to check the stability analysis for large amplitudes of diameter fluctuations, Monte Carlo simulations of a wire decay driven by surface diffusion were performed. Volume diffusion is explicitly prohibited within the simulation, i.e. atoms are not allowed to detach from the wire surface and vacancies can not be generated within the wire. Snap shots of the wire evolution can be seen in Fig.~\ref{fig:fragmentation}. The initial wire, $R_{0}=\unit{4.3}{\nano\meter}$, shows after a long incubation time statistical fluctuations, which grow faster and faster with increasing amplitude. They finally lead to the disintegration of the wire after \unit{5600}{\kilo MCS} into droplets. The average diameter of these droplets is \unit{16.4}{\nano\meter}, which is in good agreement with the prediction of the linear stability analysis, $d_{cluster}=3.78\cdot R_{0}=\unit{16}{\nano\meter}$. Note that the simulation was carried out at high temperature. The ratio between the \emph{effective} bond strength $E_{bond}$ and the thermal energy, $k_{B}T$, was set to $0.7$. An absolute value of the temperature $T$ can not be given since the value of effective Ge-Ge bond strength in SiO$_{2}$, $E_{bond}$, is unknown. It may be estimated, however, within a good
approximation from the solubility of Ge in SiO$_{2}$.

The stability analysis as well as the Monte Carlo simulations neglect the influence of grain boundaries of a polycrystalline wire. Grain boundary diffusion may provide another source of instability, if annealing is carried
out below the melting point. Then, the size of the droplets may differ from the result given above.

\section{Summary}

A novel fabrication method of buried nanowires by ion implantation has been demonstrated. Ge$^{+}$ were implanted into oxide covered Si V-grooves leading to Ge enrichment in the oxide at the V-groove bottom. There, during annealing,
the Ge forms a nanowire extending along the V-groove. Other than a similar method published recently \cite{Ishikawa:1999}, this method should work for any implanted ion species immiscible in SiO$_{2}$. In particular, ion beam synthesized Au wires, which are subsequently disintegrated in regular nanocluster chains, are of interest for the fabrication of plasmonic structures \cite{Brongersma:2000}.

The phase separation mechanisms were studied by means of kinetic 3D lattice Monte Carlo simulations. As shown, a cylindrical profile with Gaussian cross section may coalesce to a continuous nanowire of only a few nanometers
diameter. For long-lasting thermal treatment atomistic simulations revealed the pearling instability. A wire decays into equal-sized and equal-spaced droplets. An average droplet diameter of approximately twice the initial diameter of the wire, as it is predicted by linear stability analysis for a wire decay by interface diffusion, could be confirmed by Monte Carlo simulations.

Thus, nanocluster chains having a good periodicity over long distances can be fabricated by the fragmentation of nanowires using the pearling instability. Thereby, the droplet diameter is independent from material parameters and is
controlled only by the diameter of the initial wire.

\begin{acknowledgments}
We wish to acknowledge stimulating discussions with H.~A.~Atwater and
M.~L.~Brongersma related to the application of metallic nanoclusters in photonics.
\end{acknowledgments}

%\bibliography{your bib file}

\end{document}